\documentclass[aps,amsmath,amssymb,nofootinbib,twocolumn,pr]{revtex4-1}
\usepackage{bm}
\usepackage{graphicx}
\usepackage{amsmath}
\usepackage{amssymb}
\usepackage{color}
\usepackage{ulem}

\usepackage{epsfig}
\newcommand{\nix}[1]{}
\usepackage[unicode=true,colorlinks=true,citecolor=blue]{hyperref}

\begin{document}

\title{Semiclassical theory of the circular photogalvanic effect in gyrotropic systems
}

\author{L.\,E.\,Golub$^1$,  E.\,L. Ivchenko$^1$, and B.\,Spivak$^2$}
\affiliation{$^1$Ioffe Institute, 194021 St.~Petersburg, Russia}
\affiliation{$^2$Department of Physics, University of
Washington, Seattle, WA 98195, USA}

\begin{abstract}
We develop a theory of circular photogalvanic effect (CPGE) for classically high photon energies which exceed  the electron  scattering rate but are small
compared to the average electron kinetic energy.  In this frequency range one can  calculate the CPGE by using two different approaches. In the fully quantum-mechanical approach we find the photocurrent density by applying  Fermi's golden rule  for indirect intraband optical transitions with virtual intermediate states both in the conduction and valence bands. In the framework of the semiclassical approach, we apply a generalized Boltzmann equation with accounts for  the Berry-curvature induced anomalous velocity, side jumps and skew scattering. The calculation is carried out for  a wurtzite symmetry crystal. Both methods yield the same results for the CPGE current demonstrating consistency  between the two approaches and applicability of the semiclassical theory for the description of nonlinear high-frequency transport.
\end{abstract}

\maketitle

\section{Introduction}
The key signature of the Circular PhotoGalvanic Effect (CPGE) is the appearance of a photocurrent
under illumination with circularly polarized light and reversal of its direction upon inversion of the light helicity. 
Physically, the CPGE can be considered as a transformation of the
photon angular momenta into a translational motion of free charge carriers. It
is an electronic analog of mechanical systems which transmit rotatory motion
to linear one like a wheel or a screw.
In time-reversal invariant systems the circular photocurrent is nonzero for point groups that allow optical activity or gyrotropy. Among 21 crystal classes lacking inversion symmetry, only three noncentrosymmetric classes T$_d$, C$_{3h}$ and D$_{3h}$ are nongyrotropic. 
The symmetry of $C_{nv}$ 
point groups ($n=3,4,6,\infty$) allows for the CPGE analogous to the wheel effect.
The CPGE was predicted by Ivchenko and Pikus \cite{IvchPikus1978} and Belinicher~\cite{Belinich}.
It was first observed and studied in Tellurium bulk crystals by Asnin et al. \cite{Asnin}, see more references in the books [\onlinecite{SturFrid,IvchBook,GanichevPrettl_book,DyakBook}].  Sometimes the term, ``circular photocurrent'', is also referred to as ``injection current''~\cite{Sipe}.  Renewed interest  in the CPGE has been stimulated by prediction of its quantization in terms of the fundamental constants $e$, $h$ and the topological charge of Weyl nodes in the Weyl semimetals~\cite{Moore2017}, see also Refs.~\cite{Lee2017,Spivak2017,Pesin1,Golub2018,Grushin,Leppenen2019,Moore2020}.
    
The initial microscopic theories of the CPGE were based on the quantum-mechanical calculations applicable when the light  frequency $\omega$ is much larger than the electron relaxation rate $\tau^{-1}$. The goal of the work~\cite{ConMat} was to pave the way for the generalized semiclassical description 
of  the circular and linear photogalvanic effects in solids, under conditions where the electron  kinetic energy $\bar{\varepsilon}$ exceeds the photon energy $\hbar \omega$.   In this regime one can use the Boltzmann kinetic equation \cite{Niu99,Sinitsyn06,rev_MacD,rev_Niu}  with the field term containing the electric field of the electromagnetic wave.

Transport analysis of photocurrents in Ref.~[\onlinecite{ConMat}] showed  that, within the semiclassical approach, the following mechanisms can contribute to the photogalvanic effects: (i)~ skew scattering which appears   when going beyond the Born approximation~\cite{Smit56,Luttinger58,Sinitsyn}, (ii)~the electron anomalous velocity due to the Berry curvature in the electron Bloch wavefunction~\cite{ChangNiu95,Niu99,Haldane}, (iii)~the side-jump contribution to the electron velocity, and (iv)~a change of electron energy caused by the side-jump in a scattering process in the presence of an external electric field~\cite{Sinitsyn,Berger,Sinitsyn06,Sinitsyn2007}. Since the mechanism (iii) is insensitive to the light helicity, only the other three  mechanisms are relevant to the CPGE.  Soon Moore and Orenstein~\cite{Moore2010} computed the Berry-phase contribution (ii)
to helicity-dependent photocurrents in realistic circumstances, namely, for semiconductor quantum wells. This  mechanism was then studied  comprehensively for different materials and electron band models~\cite{NonlinearHall,Moore2016,Te2018,BerryCurvaturedipole2018,BerryDipole,BerryPhase,
Polini2018,BerryCurvatureDipole2019,WTe22019,SnTe,Review2019,Pesin2,BerryCurvatureDipole2020,
Wannier90}. The Berry-curvature-related circular photocurrent has been expressed in an elegant form via  the newly coined ``Berry curvature dipole''~\cite{NonlinearHall}.

In contrast to the Berry curvature-related current~(ii), the mechanism (iv) has not attracted   as much  attention although it can lead to  circular photocurrents of the same order of magnitude. Moreover,   a similar mechanism has been studied in linear and nonilinear transport,  the so-called anomalous-distribution mechanism of the anomalous Hall effect~\cite{Sinitsyn,Sinitsyn2007,Landa,NLH2,NLH_add,GlazovGolub}.

In this work, we present, within the same electron band structure model, both quantum-mechanical and semiclassical calculations of the intraband CPGE for a three-dimensional electron gas in the  frequency range
\begin{equation} \label{nonequal}
\frac{\hbar}{\tau} \ll \hbar \omega \ll \bar{\varepsilon}\:.
\end{equation} 
We will show that in this frequency range both approaches yield  the same result which indicates consistency of the semiclassical approach. We will also discuss relative roles of above-mentioned contributions to circular photocurrents.

The paper is organized as follows. In Sec.~\ref{Bandstructure} 
we present a band structure model of a wurtzite-type semiconductor used in the explicit consideration.
In Sec.~\ref{III} we perform the quantum-mechanical calculation of the circular photocurrent, and Sec.~\ref{Semicl} is devoted to the CPGE semiclassical description. The skew scattering contribution is calculated in Sec.~\ref{Skew}. Concluding remarks are given in Sec.~\ref{Concl}.

\section{Band structure model}
\label{Bandstructure}

\begin{figure}[h]
\includegraphics[width=\linewidth]{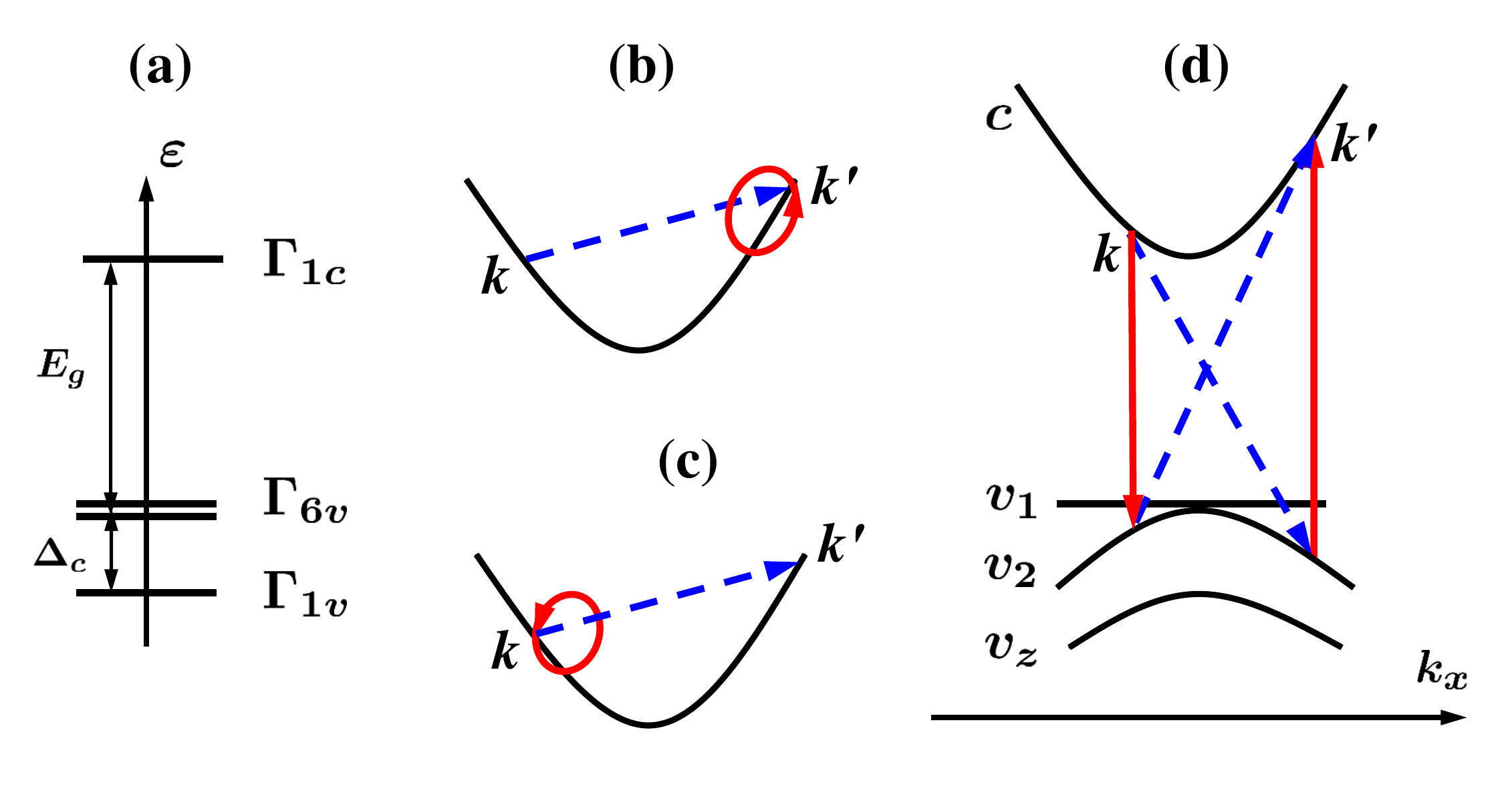}
\caption{(a): The four-band model, the electron energies at the $\Gamma$ point (${\bm k} = 0$). 
(b), (c): Indirect intraband transitions via intermediate states in the conduction band. (d): Transitions with virtual states in the valence bands.}
\label{fig}
\end{figure}

We use the spin-independent band-structure model of a wurtzite-type semiconductor
  which includes  the conduction band $\Gamma_{1c}$, and the valence bands
$\Gamma_{6v}, \Gamma_{1v}$, Fig.~\ref{fig}(a). The (real) Bloch functions at the $\Gamma$ point are labeled as $S$ ($\Gamma_{1c}$), $X, Y$ ($\Gamma_{6v}$) and $Z$ ($\Gamma_{1v}$). In this basis the Kane effective Hamiltonian is a 4$\times$4 matrix
\begin{equation} \label{Hamiltonian}
\hat{H}({\bm k})= \left[
\begin{array}{cccc}
0&{\rm i} P_{\perp} k_x &{\rm i} P_{\perp} k_y &{\rm i} P_{\parallel} k_z\\
- {\rm i} P_{\perp} k_x &-E_g&0&- {\rm i} Q k_x \\
- {\rm i} P_{\perp} k_y &0&-E_g &- {\rm i} Q k_y\\
- {\rm i} P_{\parallel} k_z&{\rm i} Q k_x& {\rm i} Q k_y& - (E_g + \Delta_c)
\end{array}
\right],
\end{equation}
where ${\bm k}$ is the electron wave vector, $E_g$ is the band gap, $\Delta_c$ is the crystal splitting of the valence band due to the uniaxial symmetry, and ${\rm i} P_{\perp}, {\rm i} P_{\parallel}, {\rm i} Q$ are interband matrix elements of the velocity operator taken between the Bloch functions at ${\bm k} = 0$ and multiplied by $\hbar$. 
This is the four-band spinless Kane Hamiltonian~\cite{IvchBook,Willatzen} generalized to account for the uniaxial anisotropy ($\Delta_c \neq 0, P_{\perp} \neq P_{\parallel}$) and broken space-inversion symmetry ($Q \neq 0$).
The electron dispersion consists of the conduction band $\varepsilon_{c,{\bm k}}$, two valence subbands $\varepsilon_{v_1,{\bm k}},\varepsilon_{v_2, {\bm k}}$ and the split-off subband $\varepsilon_{v_z,{\bm k}}$. The spin-orbit interaction is ignored, and the expressions for the electric current are just multiplied by the spin-degeneracy factor of 2. The off-diagonal components of the matrix (\ref{Hamiltonian}) describe the ${\bm k}{\bm p}$ mixing between the conduction and valence bands as well as between the valence bands $\Gamma_{6v}$ and $\Gamma_{1v}$.  For the symmetry of Bloch functions at the $\Gamma$-point we use  the notation of the point group C$_{6v}$ but the symmetry of Hamiltonian (\ref{Hamiltonian}) is higher, $C_{\infty v}$. In the chosen coordinate frame, $z$ is parallel to the principal C$_6$ axis.

The circular photocurrent is phenomenologically described by the relation $j^{\rm CPGE}_\alpha =  \gamma_{\alpha\beta} \varkappa_\beta E^2_0$, where ${\bm \varkappa = {\rm i} ({\bm e} \times {\bm e}^*)}$ is the vector chirality of a photon with ${\bm e}$ being the light polarization unit vector, $E_0$ is the light wave amplitude, and
$\hat{\bm \gamma}$ is the second rank pseudotensor~\cite{IvchBook}. For the C$_{\infty v}$ symmetry, it has only one linearly-independent component, $\gamma_{xy} = - \gamma_{yx} \equiv \gamma$:
\begin{equation} \label{gamma}
j^{\rm CPGE}_x = \gamma \varkappa_y E^2_0\:, \quad j^{\rm CPGE}_y = - \gamma \varkappa_x E^2_0\:.
\end{equation}

In the parabolic approximation the energy dispersion in the subbands $c,v_1,v_2$ and $v_z$ is given by
\begin{eqnarray}
&& \varepsilon_{ c, {\bm k} } = \frac{ \hbar^2 k_{\perp}^2 }{2 m_{c \perp} }
+ \frac{ \hbar^2 k_z^2 }{2  m_{c \parallel}} \:, \\ &&  \varepsilon_{v_1, {\bm k}} = - E_g \:, \quad \varepsilon_{v_2,{\bm k}} = - E_g - \frac{ \hbar^2 k_{\perp}^2 }{2 m_{v_2 \perp} } \:, \nonumber \\ && \varepsilon_{v_z,{\bm k}} = - (E_g + \Delta_c) -  \frac{ \hbar^2 k_{\perp}^2 }{2 m_{v_z \perp} } - \frac{ \hbar^2 k_z^2 }{2  m_{v_z \parallel}}\:,\nonumber
\end{eqnarray}
where $k^2_{\perp} = k^2_x + k^2_y$, and the inverse effective masses are
\begin{eqnarray}
&& \frac{1}{m_{c \perp}} = \frac{2 P_{\perp}^2}{\hbar^2E_g}\:, \quad \frac{1}{m_{c \parallel}} = \frac{2 P_{\parallel}^2}{\hbar^2 (E_g + \Delta_c)}\:,\\ &&  \frac{1}{m_{v_2 \perp}} = \frac{2}{\hbar^2} \left( \frac{P_{\perp}^2}{E_g} - \frac{Q^2}{\Delta_c} \right) \:, \nonumber\end{eqnarray}
\begin{eqnarray} \frac{1}{m_{v_z \perp}} = \frac{2 Q^2}{\hbar^2 \Delta_c}\:,  \quad \frac{1}{m_{v_z \parallel}} = \frac{2 P_{\parallel}^2}{\hbar^2 (E_g + \Delta_c)}\:.\nonumber
\end{eqnarray}
In the particular case
\begin{equation} \label{isotrop}
\frac{P_{\perp}^2}{E_g} = \frac{P_{\parallel}^2}{E_g + \Delta_c}
\end{equation}
the parabolic dispersion in the conduction band is isotropic: $m_{c \perp} = m_{c \parallel} \equiv m$
and \begin{equation} \label{isotrop2}
\varepsilon_{ c, {\bm k} } = \frac{ \hbar^2 k^2 }{2 m}\:.
\end{equation}
In order to simplify the computation of the photocurrent we further assume that the relation (\ref{isotrop}) is valid and use the isotropic electron dispersion in the conduction band.

The Bloch functions at ${\bm k} \neq 0$ are linear combinations of the basis functions $S,X,Y,Z$ multiplied by the exponential factor ${\rm e}^{{\rm i} {\bm k} {\bm r}}$, namely,
\[
\psi_{n, \bm k} ({\bm r}) = {\rm e}^{{\rm i} {\bm k} {\bm r}} u_{n{\bm k}}\:,\: u_{n{\bm k}} = C_s S + C_x X + C_y Y + C_z Z\:.
\]
The coefficients $ C_s, C_x , C_y$ and $C_z$  are usefully presented as a four-component column $\hat{C}_{n {\bm k}}$ which is an eigenvector of the matrix $\hat{H}({\bm k})$. 

In the following we will calculate the Berry curvature ${\bm \Omega}_{c {\bm k}}$ and the elementary shifts ${\bm r}_{c{\bm k}', c {\bm k}}$ of an electron in the ${\bm r}$-space under the scattering between the states  $(c, {\bm k})$ and $(c, {\bm k'})$. For this purpose we expand the Bloch periodic amplitudes $u_{n{\bm k}}$ in powers of ${\bm k}$ up to the second order. For the conduction band, the expansion has the form
\begin{eqnarray} \label{Blochc}
u_{c {\bm k}} &=& C^{(c)} S + \left( - {\rm i} { P_\perp \over E_g} - {P_\parallel Q k_z \over E_g E'_g} \right) (k_x X + k_y Y)\\&& + \left( - {\rm i} { P_\parallel k_z \over E'_g} + {P_\perp Q k^2_{\perp} \over E_g E'_g} \right) Z\:, \nonumber
\end{eqnarray}
where  $E'_g = E_g + \Delta_c$ and the normalization coefficient equals
\[
C^{(c)}= 1 - \frac12 \left( \frac{P_\perp^2 k^2_{\perp}}{E_g^2} +
\frac{P_{\parallel} ^2 k^2_z}{(E_g + \Delta_c)^2} \right) \:.
\]
The Bloch amplitude $u_{v_1 {\bm k}}$ is a simple linear combination of $X$ and $Y$: 
\begin{equation} \label{uv1}
u_{v_1, {\bm k}} = \frac{k_y}{k_{\perp}} X - \frac{k_x}{k_{\perp}} Y\:.
\end{equation}
Similarly to $u_{c {\bm k}}$, the amplitudes $u_{v_2 {\bm k}}$ and $u_{v_z {\bm k}}$ are linear combinations of three Bloch functions $S$, $(k_x X + k_yY)/k_{\perp}$ and $Z$:
\begin{eqnarray} \label{Blochz}
&&u_{v_2 {\bm k}} = \left( - {\rm i} { P_\perp k_{\perp} \over E_g} + {P_\parallel Q k_{\perp} k_z \over E_g \Delta_c}\right) S \\ &&+ C^{(v_2)} \left( \frac{k_x}{k_{\perp}} X +  \frac{k_y}{k_{\perp}} Y \right) + \left( {\rm i} { Q k_{\perp} \over \Delta_c} - {P_\perp P_\parallel k_{\perp} k_z \over E_g \Delta_c} \right) Z\:, \nonumber\\
&&u_{{v_z} {\bm k}} = \left( - {\rm i} { P_\parallel k_z \over E'_g} +{P_\perp Q k^2_{\perp} \over E'_g \Delta_c} \right) S \\ &&+   \left( {\rm i} { Q \over \Delta_c} + {P_\perp P_\parallel
k_z \over E'_g \Delta_c} \right)   (k_x X + k_y Y) + C^{(z)}Z\:. \nonumber
\end{eqnarray}

\section{Quantum-mechanical calculation of the circular photocurrent} \label{III}

In the quantum-mechanical approach the intraband circular photocurrent is generated under indirect optical transitions $(c {\bm k}) \to (c{\bm k}')$, with $\varepsilon_{c{\bm k}'}=\varepsilon_{c{\bm k}} + \hbar \omega$, involving  scattering by static defects and/or acoustic phonons. Here we focus on the elastic scattering processes and introduce the scattering matrix elements $U_{n'{\bm k}', n {\bm k} }$, both intraband and interband. Then the compound matrix element of the indirect light absorption has the form~\cite{Asnin,IvchBook,ST2020}
\begin{eqnarray} \label{Mcc}
&&M_{ c {\bm k}' , c {\bm k} } = \frac{ V_{c,c}({\bm k}') U_{c{\bm k}', c {\bm k} }}{ \varepsilon_{c {\bm k}'} -  \varepsilon_{ c {\bm k} } } + \frac{U_{ c {\bm k}' , c {\bm k} }  V_{c,c}({\bm k}) }{ \varepsilon_{c{\bm k}} -  \varepsilon_{ c{\bm k} } - \hbar \omega} \\ &&+ \sum\limits_{v = v_1, v_2 ,v_z} \left( \frac{ V_{c , v}({\bm k}') U_{v {\bm k}', c {\bm k} }}{ \varepsilon_{v {\bm k}'} -  \varepsilon_{c{\bm k}}  }+  \frac{ U_{c {\bm k}', v {\bm k} } V_{v , c}({\bm k}) }{ \varepsilon_{v {\bm k}} -  \varepsilon_{c {\bm k}} - \hbar \omega }\right)\:,\nonumber
\end{eqnarray}
where $V_{n',n}({\bm k})$ is the optical matrix element of the direct transition $(n {\bm k}) \to (n'{\bm k})$.

The first two terms in the rhs of Eq.~(\ref{Mcc}) describe the indirect transitions via intermediate states in the conduction band $c$ (intraband contribution), Figs.~\ref{fig}(b) and~(c), while the sum over $v$ accounts for the virtual intermediate states in the valence bands (interband contribution), Fig.~\ref{fig}(d). The electric field of the incident light is defined as
\begin{equation} \label{electricfield}
{\bm E}(t) = {\bm E} {\rm e}^{- {\rm i} \omega t} +
{\bm E}^* {\rm e}^{{\rm i} \omega t} = E_0 \left(  {\rm e}^{- {\rm i} \omega t} {\bm e} +  {\rm e}^{ {\rm i} \omega t} {\bm e}^*  \right)\:.
\end{equation}
Then the intraband contribution to (\ref{Mcc}) can be written as
\begin{equation} \label{intra}
M^{\rm intra}_{ c {\bm k}' , c {\bm k} } =  \frac{{\rm i} e E_0}{m \omega} \frac{U_{c{\bm k}', c {\bm k} }}{\omega}\left[ {\bm e} \cdot ({\bm k}' -{\bm k}) \right]\:,
\end{equation}
where $e$ is the electron charge and the effective mass  $m$ is introduced in Eq.~(\ref{isotrop2}).

Within the accuracy of our calculation the electron energy dispersion and the photon energy in the denominators of the interband contribution can be neglected, and this contribution is reduced to 
\begin{equation} \label{inter}
M^{\rm inter}_{ c {\bm k}' , c {\bm k} } = - \sum\limits_{v}  
\frac{V_{c , v} ({\bm k}')  U_{v {\bm k}', c {\bm k} } +  U_{c {\bm k}', v {\bm k} } V_{v , c}({\bm k}) }{ E^0_c -  E^0_ v  }\:,
\end{equation}
where $E^0_c, E^0_v$ are the electron energies at ${\bm k} = 0$. 

We assume that  electrons are elastically scattered by the short-range defects.  In this case the scattering matrix elements are given in the Born approximation by
\begin{equation} \label{Born}
U_{n' {\bm k}', n {\bm k} } = U_0 \langle u_{n' {\bm k}'}| u_{n{\bm k}}\rangle\:,
\end{equation} 
where $U_0$ is  real. 
Note that the perturbative expansion of the interband matrix element $U_{c {\bm k}', v {\bm k} }$ or $U_{v {\bm k}', c {\bm k} }$ starts from the first order in ${\bm k}$.

The optical matrix element is given by
\begin{equation} \label{optical}
V_{n'n}({\bm k}) = \frac{{\rm i} e E_0}{\omega}  {\bm e}\cdot{\bm v}_{n'n}({\bm k}) \:,
\end{equation}
where the velocity matrix elements are calculated according to
\begin{equation} \label{velocity}
{\bm v}_{n'n}({\bm k}) = \hat{C}^{\dag}_{n'{\bm k}} \hat{\bm v} \hat{C}_{n {\bm k}}\:, \quad \hat{\bm v} = \frac{1}{\hbar} \frac{\partial \hat{H}(\bm k)}{\partial {\bm k}}\:.
\end{equation}

Omitting tedious intermediate  transformations we arrive at the following  equation for the indirect optical matrix element
\begin{eqnarray}
&&M_{ c {\bm k}' , c {\bm k} } = {\rm i} a {\bm e}\cdot ({\bm k}'-{\bm k}) \\ && - b \left[ e_z (k_{\perp}^{\prime 2} - k_{\perp}^2 ) - {\bm e}_{\perp}\cdot \left({\bm k}'_{\perp} + {\bm k}_{\perp}\right) (k'_z - k_z) \right]\:, \nonumber
\end{eqnarray}
where
\begin{eqnarray} \label{ab}
&& a =  \frac{e E_0}{m \omega^2} U_0 \:, \quad  b =  \frac{e E_0 {\cal A} }{2 \hbar \omega} U_0\:, \\&& {\cal A} =  \frac{ 2 P_{\perp} P_{\parallel} Q}{E_g E'_g} \left( \frac{1}{E_g} + \frac{2}{E'_g} \right)\:. \label{calA}
\end{eqnarray}

The circular photocurrent is calculated according to~\cite{IvchBook}
\begin{eqnarray} \label{QMcurrent}
&&{\bm j}^{\rm CPGE} 
= 2 e n_i \frac{2\pi}{\hbar} \sum_{{\bm k}'{\bm k}} \left[ {\bm v}_{c{\bm k}'} \tau(\varepsilon_{k'}) -  {\bm v}_{c {\bm k}} \tau(\varepsilon_k)\right] 
\\ &&\times | M_{ c {\bm k}' , c {\bm k} } |^2_{\rm as} \left[ f^0(\varepsilon_k) - f^0(\varepsilon_{k'}) \right]  \delta(\varepsilon_{k'} - \varepsilon_k - \hbar \omega)\:,\nonumber \end{eqnarray}
where 
 the factor of 2 accounts for the double spin degeneracy and for brevity we set the sample volume to unity. The other notations are defined as follows: $n_i$ is the defect concentration, ${\bm v}_{c {\bm k}} = \hbar {\bm k}/m$ is the electron velocity in the conduction band, $\tau(\varepsilon_k)$ is the energy-dependent momentum scattering time, $f^0(\varepsilon_k)$ is the equilibrium electron distribution function, and we retain only the antisymmetric part of the squared matrix element
\begin{eqnarray} \label{asymm}
| M_{ c {\bm k}' , c {\bm k} } |^2_{\rm as} &=& ab \left\{ \left( k_{\perp}^{\prime 2} - k_{\perp}^2 \right) \left[ \left( {\bm k}_{\perp}' - {\bm k}_{\perp} \right) \times {\bm \varkappa} \right]_z \right. \\&&  \left. + \left( k_z' - k_z \right)^2  \left[ \left( {\bm k}_{\perp}' + {\bm k}_{\perp} \right) \times {\bm \varkappa} \right]_z \right. \nonumber \\ && \left. + 2 \left( k_z' - k_z \right) \left( {\bm k}' \times {\bm k} \right)_z \varkappa_z \right\}\:. \nonumber
\end{eqnarray} 
The last term in the rhs  of Eq.~(\ref{asymm}) is proportional to $\varkappa_z$ and does not contribute to the current in agreement with the symmetry consideration (\ref{gamma}). In the model under study the momentum relaxation rate has the form
\begin{equation}
\frac{1}{\tau(\varepsilon_k)} = \frac{\pi}{\hbar} n_i U_0^2 g(\varepsilon_k) = \frac{\sqrt{2}}{\pi} \frac{m^{3/2}}{\hbar^4}U_0^2 n_i \sqrt{\varepsilon_k}\:,
\end{equation} 
where $$g(\varepsilon_k)=2 \sum_{{\bm k}'} \delta(\varepsilon_{k'} - \varepsilon_k)\:$$ is the density of states.
Thus, the exponent of the power-law dependence $\tau (\varepsilon_k) \propto \varepsilon_k^{\nu} $ has the value
\begin{equation} \label{nu}
\nu = \frac{d \ln{\tau}}{d \ln{\varepsilon_p}} = - \frac12\:.
\end{equation}
In this paper we keep the general notation $\nu$ to show the origin of the particular contributions to the photocurrent due to the derivative $d \tau / d \varepsilon_k$. Only at the final stage we substitute the value $\nu = - 1/2$ into the equation for the photocurrent.

While performing the summation in Eq.~(\ref{QMcurrent}) we take into account the energy conservation law $\varepsilon_{k'} - \varepsilon_k = \hbar \omega$ and use the Taylor's  linear expansion
\begin{eqnarray}
\hspace{5 mm}f^0(\varepsilon_{k'}) = f^0(\varepsilon_k) + \frac{d f^0(\varepsilon_k)}{d \varepsilon_k} \hbar \omega\:, \nonumber\\
\tau(\varepsilon_{k'}) = \tau(\varepsilon_k) + \frac{d \tau(\varepsilon_k)}{d \varepsilon_k} \hbar \omega\:, \hspace{5 mm} \nonumber
\end{eqnarray}
and  the identities
\begin{eqnarray}   
&&  \frac{2 \pi}{\hbar} \sum_{{\bm k}'} \delta(\varepsilon_{k'} - \varepsilon_k) = \frac{1}{n_i U_0^2 \tau} \:, \label{sumdelta}\\
&& - 2 \sum\limits_{\bm k} \varepsilon_k \frac{d f^0}{d \varepsilon_k} = \frac32 N\:,\hspace{1 cm} \label{edfde}
\end{eqnarray}
where $N= 2\sum\limits_{{\bm k}} f^{0}(\varepsilon_{k})$ is the electron density. The final result for the CPGE parameter $\gamma$ reads
\begin{equation} \label{QMfinal}
\gamma = 2 \left( 1 + \frac{\nu}{3} \right) \frac{e^3 {\cal A} N }{\hbar^2 \omega }\:.
\end{equation}

Now we turn to the semiclassical approach and show that the  semiclassical circular photocurrent is also given by Eq.~(\ref{QMfinal}) with $\nu = - 1/2$.

\section{Semiclassical description of CPGE}
\label{Semicl}

In this section we consider  semiclassical contributions to $\gamma$  mentioned in the Introduction  and compare their sum with Eq.~(\ref{QMfinal}).
\subsection{Berry curvature dipole (BCD) contribution} \label{IIIA}
This contribution to the photocurrent is given by~\cite{ConMat}
\begin{equation} \label{BerryCurrent}
{\bm j}^{\rm BCD} = \frac{2 e^2}{\hbar} \sum\limits_{\bm k}  \left[{\bm \Omega}_{c{\bm k}} \times
{\bm E}(t) \right]\: f^{(1)}_{c{\bm k}}(t)\:,
\end{equation}
where we use the following notation. The Berry curvature ${\bm \Omega}_{c{\bm k}}$ is the curl of the Berry connection ${\bm A}_{c {\bm k}}$ (also called the Berry vector potential),
\begin{equation}
{\bm \Omega}_{c{\bm k}} =  \frac{\partial}{\partial {\bm k}} \times {\bm A}_{c {\bm k}}, \quad
{\bm A}_{c{\bm k}} = {\rm i} \left< u_{c \bm k} \Bigg| \frac{\partial
u_{c \bm k}}{\partial {\bm k}} \right>\:,
\end{equation}
and $f^{(1)}_{c{\bm k}}(t)$ is the correction to the electron distribution function linear in the electric field ${\bm E}(t)$,
\begin{equation} \label{f1t}
f^{(1)}_{c{\bm k}}(t) = {\rm e}^{- {\rm i} \omega t} f^{(1)}_{c{\bm k} \omega} +  {\rm e}^{ {\rm i} \omega t} f^{(1)*}_{c{\bm k} \omega} \:.
\end{equation}
In the framework of the classical Boltzmann equation in the relaxation-time approximation, it has the form
\begin{equation} \label{f1}
 f^{(1)}_{c{\bm k} \omega} = - e \tau_{\omega}( \varepsilon_k) E_0 ({\bm e} \cdot {\bm v}_{c{\bm k}}) \frac{d f^0(\varepsilon_k)}{d \varepsilon_k}\:,\quad \tau_{\omega} = \frac{\tau}{1 - {\rm i} \omega \tau}\:.
\end{equation}
In the limit $\omega \tau \gg 1$, see Eq.~(\ref{nonequal}), the ``complex time'' $\tau_{\omega}$ is replaced by ${\rm i}/\omega$.

One can show using Eq.~(\ref{Blochc}) for the periodic amplitude $u_{c {\bm k}}$ that the Berry curvature is given,  up to the  first order, by
\begin{equation} \label{OmegaA}
{\bm \Omega}_{c {\bm k} } = {\cal A} \left( {\bm k} \times {\bm c} \right) \:, 
\end{equation}
where ${\bm c}$ is the unit vector in the $z$ direction and  the coefficient ${\cal A}$ is defined by Eq.~(\ref{calA}). 
An important point to stress is that the Berry curvature is proportional to the same parameter ${\cal A}$ which is present in Eq.~(\ref{QMfinal}) for the quantum-mechanical photocurrent. Substituting (\ref{electricfield}), (\ref{f1}) and (\ref{OmegaA}) into Eq.~(\ref{BerryCurrent}) and summing over ${\bm k}$ we obtain
\begin{equation} \label{BCD}
\gamma^{\rm BCD} = \frac{e^3 {\cal A} N}{\hbar^2 \omega}\:.
\end{equation}
Comparing Eqs.~(\ref{QMfinal}) and (\ref{BCD}) we see that the BCD contribution represents only a part of the total CPGE current and we need to consider other mechanisms of the semiclassical circular photocurrent.
\subsection{Circular photocurrent due to the side-jump correction in the energy-conservation law} \label{IIIB}
The kinetic equation for the electron distribution function $f_{c {\bm k}}$ with allowance for the side-jumps is written as~\cite{ConMat}
\begin{equation} \label{kineticeq}
\frac{\partial f_{c {\bm k}}(t)}{\partial t} + \frac{e {\bm E}(t)}{\hbar} \frac{\partial f_{c {\bm k}}(t)}{\partial {\bm k}} =n_i \sum_{{\bm k}'} W_{{\bm k}'{\bm k}} (f_{c {\bm k}'} - f_{c {\bm k}})\:,
\end{equation}
where
\begin{eqnarray}
&& W_{{\bm k}'{\bm k}}= \frac{2 \pi}{\hbar} U_0^2 \delta[\varepsilon_{k'} - \varepsilon_k - e {\bm E}(t)\cdot {\bm r}_{{\bm k}' {\bm k}} ] \approx  W^{(0)}_{{\bm k}'{\bm k}} +  W^{(1)}_{{\bm k}'{\bm k}}\:, \nonumber\\ && W^{(0)}_{{\bm k}'{\bm k}} = \frac{2 \pi}{\hbar} U_0^2 \delta( \varepsilon_{k'} - \varepsilon_k )\:, \label{W0}\\&&W^{(1)}_{{\bm k}'{\bm k}} = \frac{2 \pi}{\hbar} e U_0^2  \left[ {\bm E}(t) \cdot {\bm r}_{{\bm k}' {\bm k}} \right] \frac{\partial \delta(\varepsilon_k - \varepsilon_{k}')}{\partial \varepsilon_k}\:, \label{W1}
\end{eqnarray}
and ${\bm r}_{c{\bm k}', c {\bm k}}$ is an elementary shift of the electron in the real space under the transition from $(c,{\bm k})$ to $(c,{\bm k}')$. Equation (\ref{kineticeq}) differs from the standard Boltzmann equation by the coordinate-shift correction to the collision integral. This correction follows from the change of the potential energy by $- e {\bm r}_{{\bm k}' {\bm k}} \cdot{\bm E}$ under the scattering in the electric field due to the side-jump in which case the energy-conservation law reads~\cite{Sinitsyn,Landa}
\[
\varepsilon_{k'} - \varepsilon_k - e {\bm r}_{{\bm k}' {\bm k}} \cdot{\bm E} = 0\:.
\]

The side-jump, or the shift in real space due to scattering, is connected with the Berry vector potential by~\cite{BelIvchStur,Sinitsyn,Twophot}
\begin{equation} \label{sp_shift}
{\bm r}_{{\bm k}' {\bm k}} =  - \frac{ {\rm Im} \left[ U_{c{\bm k}', c{\bm k}}^*
({\bm \nabla}_{{\bm k}'} +{\bm \nabla}_{\bm k}) U_{c{\bm k}', c{\bm k}}
\right]}{\left\vert U_{c{\bm k}', c{\bm k}}
\right\vert^2} +{\bm A}_{c {\bm k}'} - {\bm A}_{c {\bm k}}\:.
\end{equation}
The matrix element of the elastic scattering $U_{c{\bm k}', c{\bm k}}$ is calculated according to Eqs. (\ref{Blochc}) and (\ref{Born}) with the result
\begin{equation} \label{rk'k}
{\bm r}_{{\bm k}' {\bm k}} = \frac{\cal A}{2} \left[(k_z - k'_z)  ({ \bm k}'_{\perp} + { \bm k}_{\perp})  + \left( k_{\perp}^{\prime 2} -  k_{\perp}^2\right)  {\bm c} \right] 
\end{equation}
determined by the same parameter ${\cal A}$ as the currents in Eqs.~~(\ref{QMfinal}) and (\ref{BCD}).

The photocurrent induced by the side-jump contribution to the collision integral (SJCI) is calculated according to
\begin{equation} \label{SJCI}
{\bm j}^{\rm SJCI} = 2 e \sum_{\bm k} \frac{\hbar {\bm k}}{m} g_{\bm k}\:,
\end{equation}
where $g_{\bm k}$ is the steady-state correction to the electron distribution function proportional to $E_0^2$. It consists of two contributions $g^a_{\bm k}$ and $g^b_{\bm k}$. 

\subsubsection{Field effect followed by SJCI}
The contribution $g^a_{\bm k}$ is obtained by substituting the correction $f^{(1)}_{c{\bm k}}(t)$ induced by the electric field, see Eqs.~(\ref{f1t}) and (\ref{f1}), into the rhs of Eq.~(\ref{kineticeq}) and averaging the product $$W^{(1)}_{{\bm k}'{\bm k}}[f^{(1)}_{c{\bm k}'}(t) - f^{(1)}_{c{\bm k}}(t)]$$ over time. This leads to 
\begin{eqnarray} \label{xia}
&&g^a_{\bm k} = \tau n_i \sum_{{\bm k}'} \overline{W^{(1)}_{{\bm k}'{\bm k}}[f^{(1)}_{c{\bm k}'}(t) - f^{(1)}_{c{\bm k}}(t)]} = - e \tau n_i U_0^2 \\ &&\times \frac{2 \pi}{\hbar} \sum_{{\bm k}'}\overline{\left[ {\bm E}(t) \cdot {\bm r}_{{\bm k}' {\bm k}} \right][f^{(1)}_{c{\bm k}}(t) - f^{(1)}_{c{\bm k}'}(t)]}\frac{\partial \delta(\varepsilon_k - \varepsilon_{k}')}{\partial \varepsilon_k}\:, \nonumber
\end{eqnarray}
 where the overline denotes a time average.
We substitute $f^{(1)}_{c{\bm k}}(t)$ from (\ref{f1t}), (\ref{f1}) and ${\bm r}_{{\bm k}' {\bm k}}$
from (\ref{rk'k}) into Eq.~(\ref{xia}), average over the time and the direction of ${\bm k}'$ and then substitute $g^a_{\bm k}$ into Eq.~(\ref{SJCI}), average over the direction of ${\bm k}$ and apply the identity (\ref{sumdelta}) and the sum rule
\begin{eqnarray}
&&\hspace{1.2 cm}\sum_{\bm k} F(\varepsilon_k )\frac{\partial \delta(\varepsilon_k - \varepsilon_{k}')}{\partial \varepsilon_k}\\ && = - \nonumber \sum_{\bm k} \delta(\varepsilon_k - \varepsilon_{k}') \frac{1}{\sqrt{\varepsilon_k}} \frac{\partial}{\partial \varepsilon_k} \left[  \sqrt{\varepsilon_k} F(\varepsilon_k ) \right]\:, \nonumber
\end{eqnarray}
 which is derived  by integrating  by parts and   valid for an arbitrary differentiable function $F(\varepsilon_k )$. This procedure allows us to reduce the contribution to the CPGE parameter $\gamma$ to the sum
\[
\gamma^{{\rm SJCI}a} = - \frac{4 e^3 {\cal A}}{9 \hbar^2\omega} \sum_{\bm k}  \left[ 2 (3 + \nu) \varepsilon_k  \frac{d f^0(\varepsilon_{k})}{d \varepsilon_{k}} +  \varepsilon_k^2 \frac{d^2 f^0(\varepsilon_{k})}{d \varepsilon_{k}^2}  \right].
\]
Taking into account Eq.~(\ref{edfde}) and the similar equation
\begin{equation}
2 \sum_{\bm k}  \varepsilon_k^2  \frac{d^2 f^0(\varepsilon_{k})}{d \varepsilon_{k}^2} = \frac{15}{4} N
\end{equation}
we arrive at
\begin{equation} \label{SJCIa}
\gamma^{{\rm SJCI}a} =  \frac{ e^3 {\cal A}N}{\hbar^2\omega}\frac{7 + 4 \nu}{6}\:. 
\end{equation}
\subsubsection{ SJCI followed by  the field effect}
The correction $g^b_{\bm k}$ is obtained by first solving the kinetic equation without the field term but with the inhomogeneous term 
$$ 
n_i \sum\limits_{{\bm k}'} W^{(1)}_{{\bm k}'{\bm k}} \left[ f^0(\varepsilon_{k'}) - f^0(\varepsilon_k)\right]
$$ 
proportional to $E_0$. In the limit $\omega \tau \gg 1$, the solution reads
\begin{equation} 
\tilde{f}^{(1)}_{c {\bm k}}(t) ={\rm i} \frac{n_i}{\omega} \sum_{{\bm k}'} W^{(1)}_{{\bm k}'{\bm k}} \left[ f^0(\varepsilon_{k'}) - f^0(\varepsilon_k) \right]\:.
\end{equation}
Secondly, the solution $\tilde{f}^{(1)}_{c {\bm k}}$ is substituted into the field term of the kinetic equation. As a result we have
\begin{equation} \label{xib}
g^b_{\bm k} = - \tau \overline{\frac{e {\bm E}(t)}{\hbar} \frac{\partial \tilde{f}^{(1)}_{c {\bm k}}(t)}{\partial {\bm k}}}\:.
\end{equation}
Next we substitute (\ref{xib}) into Eq.~ (\ref{SJCI}) and integrate by parts to get
\begin{equation}
j_{\alpha}^{{\rm SJCI}b} = \frac{2 e^2}{m} \sum_{\bm k}  \left[  \overline{{\bm E}(t) \tilde{f}^{(1)}_{c {\bm k}}(t)} \right] \frac{\partial [k_{\alpha} \tau(\varepsilon_k)]}{\partial {\bm k}}\:.
\end{equation}
The function in the sum averaged over  time and  the  direction of ${\bm k}$ can be reduced to
\[
({\bm c} \times  {\bm \varkappa}) E^2_0  \frac{2 \nu e m {\cal A}}{9 \hbar^2\omega} \frac{\partial f^0(\varepsilon_k) }{\partial \varepsilon_k} \varepsilon_k\:,
\]
which finally leads to
\begin{equation} \label{SJCIb}
\gamma^{{\rm SJCI}b} = \frac{\nu}{3} \frac{e^3 {\cal A} N}{\hbar^2 \omega}  \:.
\end{equation}

The sum of contributions to $\gamma$ from the Berry curvature dipole, Eq.~(\ref{BCD}), and the side jumps in the collision integral, Eqs. (\ref{SJCIa}) and (\ref{SJCIb}), is given by
\[
\gamma^{\rm CPGE} = \left( 1 + \frac{7 + 4 \nu}{6} + \frac{\nu}{3} \right)  \frac{e^3 {\cal A} N}{\hbar^2 \omega} = \left( \frac{13}{6} + \nu \right)  \frac{e^3 {\cal A} N}{\hbar^2 \omega}\:.
\]
In the model under consideration this equation is valid only for $\nu = - 1/2$ and reduces to
\begin{equation} \label{semigamma}
\gamma^{\rm CPGE} = \frac{5}{3}  \frac{e^3 {\cal A} N}{\hbar^2 \omega} \:.
\end{equation} 
The same value of $\gamma$ follows from the quantum-mechanical result, Eq.~(\ref{QMfinal}), at $\nu = - 1/2$.

\section{The circular photocurrent due to  the skew scattering} \label{Skew}
The skew-scattering mechanism is related to the antisymmetric part  of the Boltzmann
equation collision term arising beyond the Born approximation. 
Neglecting here the side-jump term $W_{{\bm k}' {\bm k}}^{(1)}$ we can  write the probability rate 
in noncentrosymmetric solids as
\[
W_{ {\bm k}' {\bm k} } = W_{ {\bm k}' {\bm k} }^{(0)} + W_{ {\bm k}'{\bm k}}^{(a)}\:,
\]
where the main term is defined in Eq.~(\ref{W0}) and the antisymmetric term possesses the properties
\[
W_{ {\bm k}'{\bm k}}^{(a)} = - W_{- {\bm k}', - {\bm k}}^{(a)} = - W_{ {\bm k}{\bm k}'}^{(a)}\:.
\]
With allowance for this term the rhs of the kinetic equation (\ref{kineticeq}) should be rewritten as
\[
I\{f\} = n_i \sum_{{\bm k}'} \left( W_{{\bm k}{\bm k}'} f_{c {\bm k}'} -  W_{{\bm k}'{\bm k}}  f_{c {\bm k}}\right)\:.
\]
Its antisymmetric part can be simplified to
\begin{eqnarray} \label{Iaf}
I_a\{f\} &=& n_i \sum_{{\bm k}'}  W_{{\bm k}{\bm k}'}^{(a)} \left( f_{c {\bm k}'} +  f_{c {\bm k}}\right)\\ &=& n_i \sum_{{\bm k}'}  W_{{\bm k}{\bm k}'}^{(a)} f_{c {\bm k}'} \:,\nonumber
\end{eqnarray} 
where the sum property $\sum\limits_{{\bm k}'} W_{ {\bm k} {\bm k}' }^{(a)} = 0$ is taken into account.

 To the third order in $U$  the antisymmetric part of the elastic scattering probability is given by \cite{SturFrid,Sinitsyn06,Sturman1984,LPGE_TI}
\begin{eqnarray} \label{WU3}
&&W_{ {\bm k}{\bm k}'}^{(a)} =  \frac{(2 \pi)^2}{\hbar} \sum\limits_{{\bm k}''} U_{c {\bm k}, c {\bm k}''} U_{c {\bm k}'', c {\bm k}'} U_{c {\bm k}', c {\bm k}}\\ &&\hspace{1.5 cm}\times\ \delta (\varepsilon_{ck'}-\varepsilon_{ck}) \delta (\varepsilon_{ck''}-\varepsilon_{ck}) \:.\nonumber
\end{eqnarray}
In the band model (\ref{Hamiltonian}) we find the following expression
\begin{eqnarray}
&& \hspace{1 cm}W_{ {\bm k}{\bm k}'}^{(a)} =  \pi W_{ {\bm k} {\bm k}' }^{(0)} g(\varepsilon_k) \frac{ P_{\perp} P_{\parallel} Q U_0 }{ E_g E'_g } \\ && \times \left[ \left( \frac{ {\bm k}'_{\perp} \cdot {\bm k}_{\perp} }{E_g} - \frac23 \frac{k^2}{E'_g} \right) (k'_z - k_z)  + \frac{k'_z k^2_{\perp}- k_z k^{\prime 2}_{\perp}}{E'_g} \right]\:. \nonumber
\end{eqnarray}

The PGE current is given by Eq.~(\ref{SJCI}) where $g_{\bm k}$ is replaced by the quadratic-in-$E_0$ correction, $f^{(2a)}_{\bm k}$, averaged over time. It is calculated in three steps. 
 First, we find the first-order correction $f^{(1)}_{c{\bm k}}$, see Eq.~(\ref{f1t}). Then we substitute it into the collision integral and determine a correction, $f^{(1a)}_{c{\bm k}}$, to the distribution function induced by the antisymmetric part (\ref{Iaf})
\[
f^{(1a)}_{c {\bm k}\omega} = \tau_{\omega} I_a\{f^{(1)} \}\:.
\]
 Lastly, we find the desired second-order correction
\[
f^{(2a)}_{c{\bm k}}  = - \tau \frac{e {\bm E}}{\hbar}\frac{\partial f^{(1a)}_{c {\bm k}}}{\partial {\bm k}}\:.
\]

The photocurrent can be rewritten as
\begin{eqnarray}
&&{\bm j} = 2 e \sum\limits_{\bm k} {\bm v}_{c {\bm k}} f^{(2a)}_{c{\bm k}} =  \frac{2e^3 E_0^2}{\hbar} \sum\limits_{\bm k} {\bm v}_{c {\bm k}} \tau \\ && \times
\left( {\bm e} \cdot \frac{ \partial }{ \partial {\bm k} } \right) \left[ \tau^2_{\omega} I_a \left\{ {\bm e}^* \cdot {\bm v}_{c{\bm k}' } \frac{ \partial f^{0}(\varepsilon_{k'})}{\partial \varepsilon_{k'}}\right\} \right] + \mbox{c.c.}\nonumber
\end{eqnarray}
In the limit $\omega \tau \to \infty$ when $\tau^2_{\omega} \to - \omega^{-2}$ the circular photocurrent vanishes and one needs to take into account the next-order correction in the expansion
\begin{equation} \label{tau2}
\tau^2_{\omega} \approx - \frac{1}{\omega^2} \left( 1 + \frac{2 {\rm i}}{\omega \tau} \right) \:.
\end{equation}
Omitting the details, we present the result for the degenerate electron gas
\begin{equation} \label{finskew}
\gamma^{\rm skew} = - \frac{8 \pi \nu}{15} \frac{e^3{\cal A}' N}{\hbar^2 \omega} \frac{\varepsilon_F g(\varepsilon_F) U_0}{\hbar \omega} \frac{1}{\omega \tau}\:,
\end{equation} 
\\where $\varepsilon_F$ is the Fermi energy, $\tau = \tau(\varepsilon_F)$,
\[
{\cal A}' =  \frac{ P_{\perp} P_{\parallel} Q}{ E_g E'_g }\left( \frac{1}{E_g} + \frac23 \frac{1}{E'_g} \right)\:,
\]
and the dimensionless parameter $g(\varepsilon_F) U_0$ is assumed to be small which is the criterion for validity of the expansion of the collision term in powers of $U$ leading to Eq.~(\ref{WU3}).
Note that the ratio of $\gamma^{\rm skew}$ to the $\gamma^{\rm CPGE}$ value of Eq.~\eqref{semigamma} is a product of a large parameter $\varepsilon_F/\hbar\omega$ and two small parameters, $g(\varepsilon_F) U_0$ and $(\omega \tau)^{-1}$.
It is interesting to mention that the dependence $\gamma^{\rm skew}\propto 1/\omega^3$ is typical for the skew-scattering induced CPGE at high frequencies~\cite{Magarill_Entin,Golub_twisted}.
It should also be stressed that the CPGE photocurrent~(\ref{finskew})  arises due to the $(\omega \tau)^{-1}$ correction in Eq.~(\ref{tau2}). This means that the result (\ref{finskew}) cannot be obtained in the quantum-mechanical approach used in Sect. \ref{III} in the limit $\omega \tau \to \infty$, and one needs to apply the Green's function diagram technique which is beyond of this study.

According to Eq.~(\ref{gamma}) the macroscopic parameter $\gamma$ relates the variables ${\bm j}$ and ${\bm \varkappa} E_0^2$  which both change the sign under the time reversal ${\cal T}$. This explains why the
 expression $\gamma^{\rm CPGE}$ in Eq.~(\ref{semigamma}) does not contain dissipative parameters. The expression for $\gamma^{\rm skew}$  agrees with  the time inversion symmetry considerations 
 because the presence of the dissipative factor $\tau^{-1}$ is compensated by an extra $\delta$-function in Eq.~(\ref{WU3}).

The anomalous Hall coefficient also allows for a fourth-order correction, $U^4$,  to the antisymmetric part of the scattering rate \cite{Sinitsyn}. An  estimate for $\omega \tau \gg 1$ shows that this skew contribution to $\gamma$ is smaller than that in Eq.~(\ref{semigamma}) by $(\omega \tau)^2$ and can be neglected.

\section{Concluding remarks}
\label{Concl}

The semiclassical approach in the nonlinear transport is based on the two conditions: (i) the photon energy $\hbar \omega$ is much smaller than the typical electron energy $\bar{\varepsilon}$, and (ii)
in the kinetic equation for the free carrier distribution function, the standard or generalized Boltzmann equation, the  field term  
\begin{equation}
\dot{\bm k}_c \frac{\partial f_{ c {\bm k} } }{\partial {\bm k} }
\end{equation}
contains the electric and magnetic fields of the light wave as follows
\[
\hbar \dot{\bm k}_c = e \left( {\bm E} + \frac{\dot{\bm r}_c}{c}\times {\bm B} \right)\:,
\]
where $\dot{\bm r}_c$ is the generalized semiclassical velocity \cite{Niu99}
\[
\dot{\bm r}_c = \frac{\hbar {\bm k}}{m} + \frac{e}{\hbar}\ {\bm \Omega}_{c{\bm k}} \times {\bm E}\:.
\]
Both the previous and present study show that the challenge in calculating a second-order response to 
electromagnetic waves is related to the choice of the collision term. The latter has corrections due to the side-jump, Sect. \ref{IIIB}, and skew, Sect. \ref{Skew}, scatterings.

The alternative quantum-mechanical approach is to calculate the optical transition matrix elements linear in the electric or magnetic field of the light wave.  Microscopically, the total photocurrent density is generally a sum of two different contributions, ballistic and shift currents, ${\bm j} = {\bm j}_b + {\bm j}_{sh}$~\cite{Sturman_2020}. To calculate the ballistic photocurrent one 
needs to solve the Boltzmann equation where the electromagnetic field is present in the generation rate proportional to the squared transition matrix element, 
calculated with allowance for  the asymmetry of  the electron-photon and/or electron-phonon or electron-impurity interaction. The shift photocurrent is determined by a product of the light-induced quantum transition probability rate and the elementary displacement (side-jump) ${\bm r}_{m {\bm k}', n {\bm k}}$ of a free carrier at the moment of the transition.

In the present work we have  aimed to reconcile the two approaches.
 It is possible to do in the frequency range~(\ref{nonequal}). Firstly, the agreement between  
  results~\eqref{QMfinal} and~\eqref{semigamma} for the circular photocurrent confirms the validity of the theory. 
Secondly, we have shown that the agreement is achieved  only 
 if, in the semiclassical approach, 
one takes into account  the Berry curvature dipole mechanism and the contribution from the field-induced side-jump correction to the transferred electron  energy.

\acknowledgments

The work of L.E.G. was supported by the Foundation for the Advancement of Theoretical Physics and Mathematics ``BASIS''.


\begin{thebibliography}{80}
\bibitem{IvchPikus1978} E.L. Ivchenko and G.E. Pikus, Pis'ma Zh. Eksp. Teor. Fiz. {\bf 27}, 640 (1978) [JETP Lett. {\bf 27}, 604 (1978)].
\bibitem{Belinich} V.I. Belinicher, Phys. Lett. A {\bf 66}, 213 (1978).
\bibitem{Asnin} V.M. Asnin, A.A. Bakun, A.M. Danishevskii, E.L. Ivchenko, G.E. Pikus,
A.A. Rogachev, Pis'ma Zh. Eksp. Teor. Fiz. {\bf 28}, 80 (1978) [JETP Lett. {\bf 28},
74 (1978)].
\bibitem{SturFrid} B.I. Sturman and V.M. Fridkin, \textit{The Photovoltaic and Photorefractive Effects
in Non-Centrosymmetric Materials} (Gordon and Breach Science Publishers, New York, 1992).
\bibitem{IvchBook} E.L. Ivchenko, \textit{Optical Spectroscopy of Semiconductor Nanostructures} (Alpha
Science Int., Harrow, UK, 2005).
\bibitem{GanichevPrettl_book}S.~D. Ganichev and W.~Prettl, \textit{Intense Terahertz Excitation of
Semiconductors} (Oxford Univ. Press, Oxford, 2006).
\bibitem{DyakBook} E.L. Ivchenko and S.D. Ganichev, Spin Photogalvanics
in \textit{Spin Physics in Semiconductors}, ed. M.I.~Dyakonov (Springer Verlag, 2016, second edition, extended), Chap.~9.
\bibitem{Sipe} J.E. Sipe and A.I. Shkrebtii, 
Phys. Rev. B {\bf 61}, 5337 (2000).
\bibitem{Moore2017} F. de Juan, A. G. Grushin, T. Morimoto, and J. E. Moore,
Nat. Commun. {\bf 8}, 15995 (2017).
\bibitem{Lee2017}C.-K. Chan, N. H. Lindner, G. Refael, and P. A. Lee, 
Phys. Rev. B {\bf 95}, 041104 (2017). 
\bibitem{Spivak2017}L. E. Golub, E. L. Ivchenko, and B. Z. Spivak, JETP
Lett. {\bf 105}, 782 (2017).

\bibitem{Pesin1}
E. J. K\"onig, H.-Y. Xie, D. A. Pesin, and A. Levchenko,
Phys. Rev. B {\bf 96}, 075123 (2017).


\bibitem{Golub2018}L. E. Golub and E. L. Ivchenko, Phys. Rev. B {\bf 98}, 075305 (2018).
\bibitem{Grushin} F. Flicker, F. de Juan, B. Bradlyn, T. Morimoto, M.G. Vergniory, and A.G. Grushin,
Phys. Rev. B {\bf 98}, 155145 (2018).
\bibitem{Leppenen2019}N. V. Leppenen, E. L. Ivchenko, and L. E. Golub, JETP {\bf 129}, 139 (2019).
\bibitem{Moore2020} A. Avdoshkin, V. Kozii, and J.E. Moore, 
Phys. Rev. Lett. {\bf 124}, 196603 (2020).
\bibitem{ConMat} E. Deyo, L. E. Golub, E. L. Ivchenko, and B. Spivak,
arXiv:0904.1917  [cond-mat.mes-hall]  13 Apr 2009.
\bibitem{Niu99} G. Sundaram and Q. Niu, Phys. Rev. B \textbf{59}, 14915 (1999).
\bibitem{Sinitsyn06} N.\,A. Sinitsyn, Q. Niu, and A.\,H. MacDonald, Phys. Rev. B \textbf{73},
075318 (2006).
\bibitem{rev_MacD}Naoto Nagaosa, Jairo Sinova, Shigeki Onoda, A. H. MacDonald, and N. P. Ong,
Rev. Mod. Phys. {\bf 82}, 1539 (2010).
\bibitem{rev_Niu}Di Xiao, Ming-Che Chang, and Qian Niu,
Rev. Mod. Phys. {\bf 82}, 1959 (2010).
\bibitem{Smit56} J. Smit, Physica \textbf{21}, 877 (1956); Physica \textbf{24},
29 (1958).
\bibitem{Luttinger58} J.\,M. Luttinger, Phys. Rev. \textbf{112}, 739 (1958).
\bibitem{Sinitsyn} N.\,A.~Sinitsyn, J. Phys.: Condens. Matter \textbf{20}, 023201 (2008).
\bibitem{ChangNiu95} M.-C. Chang and Q. Niu, Phys. Rev. Lett.
{\bf 75}, 1348 (1995).
\bibitem{Haldane} F.\,D.\,M. Haldane, Phys. Rev. Lett. \textbf{93}, 206602
(2004).
\bibitem{Berger} L. Berger, Phys. Rev. B \textbf{2}, 4559 (1970).
\bibitem{Sinitsyn2007} N.A. Sinitsyn, A.H. MacDonald, T. Jungwirth, V.K. Dugaev, and J. Sinova, 
Phys. Rev. B \textbf{75}, 045315 (2007).
\bibitem{Moore2010} J.E. Moore and J. Orenstein, 
Phys. Rev. Lett. {\bf 105}, 026805 (2010).
\bibitem{NonlinearHall} I. Sodemann and L. Fu, 
Phys. Rev. Lett. {\bf 115}, 216806 (2015).
\bibitem{Moore2016} T. Morimoto, Shudan Zhong, J. Orenstein, and J.E. Moore, 
Phys. Rev. B {\bf 94}, 245121 (2016).

\bibitem{Te2018} S.S. Tsirkin, P.A. Puente, and I. Souza, 
Phys. Rev. B {\bf 97}, 035158 (2018).
\bibitem{BerryCurvaturedipole2018} Yang Zhang, Yan Sun, and Binghai Yan, 
Phys. Rev. B {\bf 97}, 041101 (2018).
\bibitem{Polini2018} H. Rostami and M. Polini, 
Phys. Rev. B {\bf 97}, 195151 (2018).
\bibitem{BerryDipole} J.I. Facio, D. Efremov, K. Koepernik, Jhih-Shih You, I. Sodemann, and J. van den Brink, 
Phys. Rev. Lett. {\bf 121}, 246403 (2018).
\bibitem{BerryPhase} Yang Zhang, J. van den Brink, C. Felser, and Binghai Yan, 
2D Mater. {\bf 5}, 044001 (2018).
\bibitem{BerryCurvatureDipole2019} R. Battilomo, N. Scopigno, and C. Ortix, 
Phys. Rev. Lett. {\bf 123}, 196403 (2019).
\bibitem{WTe22019} Hua Wang and Xiaofeng Qian, 
npj Computational Materials {\bf 5}, 119 (2019).
\bibitem{SnTe} J. Kim, K.-W. Kim, D. Shin, S.-H. Lee, J. Sinova, Noejung Park, and H. Jin, 
Nature Communications {\bf 10}, 3965 (2019).
\bibitem{Review2019} Yang Gao, 
Front. Phys. {\bf 14}, 33404 (2019).

\bibitem{Pesin2}
E. J. K\"onig, M. Dzero, A. Levchenko, and D. A. Pesin,
Phys. Rev. B {\bf 99}, 155404 (2019).

\bibitem{BerryCurvatureDipole2020} B.T. Zhou, Cheng-Ping Zhang, and K.T. Law, 
Phys. Rev. Applied {\bf 13}, 024053 (2020).
\bibitem{Wannier90} G. Pizzi, V. Vitale, R. Arita et al., 
J. Phys.: Condens. Matter {\bf 32}, 165902 (2020). 
\bibitem{Landa} Yu. B. Lyanda-Geller, 
JETP Lett. {\bf 46}, 489  (1987).
\bibitem{NLH2} Z. Z. Du, C. M. Wang, Shuai Li, Hai-Zhou Lu and X. C. Xie, 
Nat. Commun. {\bf 10}, 3047 (2019).
\bibitem{NLH_add} 
 Cong Xiao, Z. Z. Du, and Qian Niu,
Phys. Rev. B {\bf 100}, 165422 (2019).

\bibitem{GlazovGolub} M. M. Glazov and L. E. Golub, 
arXiv:2004.05091 [cond-mat.mes-hall] 10 Apr 2020.

\bibitem{Willatzen} J. Gravesen and M. Willatzen,  
Phys. Status Solidi RRL 1800305 (2018).


\bibitem{ST2020}
S. Hubmann, G. V. Budkin, M. Otteneder, D. But, D. Sacr\'e, I. Yahniuk, K. Diendorfer, V. V. Bel'kov, D. A. Kozlov, N. N. Mikhailov, S. A. Dvoretsky, V. S. Varavin, V. G. Remesnik, S. A. Tarasenko, W. Knap, and S. D. Ganichev,
Phys. Rev. Materials {\bf 4}, 043607 (2020).


\bibitem{BelIvchStur} V.\,I. Belinicher, E.\,L.~Ivchenko, and B.\,I.~Sturman, Zh. Eksp. Teor.
Fiz. {\bf 83}, 649 (1982) [Sov. Phys. JETP {\bf 56}, 359 (1982)].
\bibitem{Twophot} L. E. Golub and E. L. Ivchenko, Zh. Exp. Theor. Fiz. {\bf 139}, 175 (2011) [JETP {\bf 112}, 152 (2011)].
\bibitem{Sturman1984} B.I. Sturman, Usp. Fiz. Nauk {\bf 144}, 497 (1984) [Sov. Phys. Usp. {\bf 27}, 881 (1984)].
\bibitem{LPGE_TI} P. Olbrich, L. E. Golub, T. Herrmann, S. N. Danilov, H. Plank, V. V. Bel'kov, G. Mussler, Ch. Weyrich, C. M. Schneider, J. Kampmeier, D. Gr\"utzmacher, L. Plucinski, M. Eschbach, and S. D. Ganichev, 
Phys. Rev. Lett. {\bf 113}, 096601 (2014). 

\bibitem{Magarill_Entin} M. V. Entin and L. I. Magarill,
Phys. Rev. B {\bf 73}, 205206 (2006).
\bibitem{Golub_twisted} M. Otteneder, S. Hubmann, X. Lu, D. Kozlov, L. E. Golub, K. Watanabe, T. Taniguchi, D. K. Efetov, and S. D. Ganichev, 	
arXiv:2006.08324 [cond-mat.mes-hall] 16~June~2020.

\bibitem{Sturman_2020} B. I. Sturman, 
Phys. Usp. {\bf 63}, 407 (2020).
\end{thebibliography}
\end{document}